\documentclass[epj]{svjour}

\usepackage{graphicx}
\usepackage{dcolumn}
\usepackage{amsmath}
\usepackage{array}
\usepackage{bm}
\usepackage{amssymb}
\usepackage{amsfonts}

\usepackage{multirow} 

\usepackage[noautoscale,vcentermath]{youngtab}
\begin{document}

\title{A minimal quasiparticle approach for the QGP and its large-$N_c$ limits}

\author{Fabien Buisseret\thanks{F.R.S.-FNRS Postdoctoral researcher; fabien.buisseret@umons.ac.be}, Gwendolyn Lacroix
%
}                     
\institute{Service de Physique Nucl\'{e}aire et Subnucl\'eaire,
Universit\'{e} de Mons -- UMONS, Place du Parc 20, 7000 Mons, Belgium}

\date{\today}

\abstract{
We propose a quasiparticle approach allowing to compute the equation of state of a generic gauge theory with gauge group SU($N_c$) and quarks in an arbitrary representation. Our formalism relies on the thermal quasiparticle masses (quarks and gluons) computed from perturbative techniques, in which the standard two-loop running coupling constant is used. Our model is minimal in the sense that we do not allow any extra ansatz concerning the temperature-dependence of the running coupling. We first show that it is able to reproduce the most recent equations of state computed on the lattice for temperatures typically higher than 2 $T_c$. Well above $T_c$ indeed, an ideal gas framework with thermal masses is indeed expected to be relevant. Then we study the accuracy of various inequivalent large-$N_c$ limits concerning the description of the QCD results, as well as the equivalence between the QCD$_{AS}$ limit and the ${\cal N}=1$ SUSY Yang-Mills theory. 
}

\PACS{
      {12.38.Mh}{} \and
      {11.15.Pg}{}
}      

\maketitle

\section{Introduction}

More than two decades after pioneering works~\cite{first}, the phenomenology related to the deconfined phase of QCD, \textit{i.e.} the celebrated quark-gluon plasma (QGP), is still a matter of intense investigations, both experimentally and theoretically. On the experimental side, the QCD matter is or will be studied in heavy-ion collisions at RHIC, SPS, FAIR, and the LHC. We refer the interested reader to the reviews~\cite{expe} for a survey of the main experimental issues in QGP physics. 

On the theoretical side, the study of gauge theories at finite temperatures, among which QCD plays a central role, deserves much interest since it is both a challenging problem in itself and possibly applicable to current experimental results. In the present study, we will focus on the two following approaches: lattice QCD and quasiparticle models. The lattice formulation of QCD allows to extract nonperturbative informations from QCD through numerical computations, and lattice data are often used to fit phenomenological model's parameters when no experimental result is available. An observable that can be accurately computed on the lattice and by resorting to effective approaches is the QCD equation of state (EoS), \textit{i.e.} the evolution of various thermodynamical quantities \textit{versus} the temperature. Although not directly measurable from experiments, the EoS is of crucial importance to understand the phase diagram of QCD. One can for example read the reviews~\cite{reveos} for complete discussions of the various EoS computed so far: with or without quark flavors, at zero chemical potential or not, \textit{etc}. More results can be found for example in~\cite{panero,detar}. 

Starting from a more phenomenological point of view, quasiparticle models are based on the assumption that the QGP can be seen as a gas of deconfined quarks and gluons\footnote{Remark that a quasiparticle description of hypothetical quark stars has been proposed historically as an application of the quark hypothesis, see \cite{itoh}.}. Pioneering works~\cite{qp1,qp2,levai} have shown that the condition for quasiparticle models to reproduce the various EoS computed in lattice QCD above the critical temperature $T_c$ is that the quasiparticles gain thermal masses. Those \textit{ad hoc} masses had to be fitted on the lattice data. More recently~\cite{str1,blaiz}, Hard-Thermal-Loop (HTL) techniques have been used to compute analytically the behavior of thermal masses at leading and next-to-leading order in the coupling $g$ from QCD. As it will be recalled in Sec.~\ref{qpmodel}, such HTL formulas are of interest to build a more firmly-based quasiparticle formalism, see \textit{e.g.}~\cite{qpmodel1}. Our aim is to use a formalism as simple and general as possible so that we can compare it to a maximal number of recent lattice results. This will be done in Sec.~\ref{glupla} for a pure Yang-Mills plasma and in Sec.~\ref{addqf} for a general QGP, with light (and heavy) quark flavors, and a nonzero chemical potential. Moreover, we will not explicitly particularize our formulas to QCD as usually done: Taking into account a SU($N_c$) gauge group and quarks in an arbitrary representation of that group, we will be able to study various large-$N_c$ limits of our model and to check their accuracy to reproduce the QCD results. To our knowledge, that problem has not been addressed yet in quasiparticle models and is studied in Sec.~\ref{lnclimits}, as well as the case of the ${\cal N}=1$ SUSY Yang-Mills EoS, that will also be computed in that section. Notice that the question of knowing ``how far $N_c=3$ is from $\infty$" is still widely studied nowadays (see \textit{e.g.}~\cite{meyer}), and has gained a renewed interest, in particular thanks to AdS/CFT-inspired approaches of QCD; one can quote for example the review~\cite{adsrev}. After some comments on heavy mesons, we summarize our results in Sec.~\ref{conclu}. 

\section{Quasiparticle model}\label{qpmodel}

Let us begin by recalling the results that will be used in the present study. It is worth saying that those formulas are not new in themselves: they have been obtained in previous studies by using perturbative methods. The peculiarity of the present work is rather to apply them without introducing phenomenological modifications, with one set of parameters for all the possible cases considered.

\subsection{Analytic results}
First of all, the two-loop running constant reads~\cite{g2r}
\begin{equation}\label{g2}
g^2=\frac{8\pi^2}{\beta_0
\ln\left(\frac{T}{\Lambda}\right)}\left\{1-\frac{\beta_1}{2\beta_0^2}\frac{\ln\left[2\ln\left(\frac{T}{\Lambda}\right)\right]}{\ln\left(\frac{T}{\Lambda}\right)}\right\},
\end{equation}
where the temperature, $T$, has been chosen as the typical energy scale of the system, while $\Lambda$ (or $\Lambda_{QCD}$) is the free parameter of the model. $\beta_0$ and $\beta_1$ are the coefficients of the two-loop $\beta$-function, reading, for a theory with gauge group SU($N_c$) and $n_f$ light quark flavors in the representation $R$ of that group,
\begin{eqnarray}\label{b01}
\beta_0&=&\frac{11}{3}N_c-\frac{4}{3}{\rm T}_R\, n_f,\nonumber\\ \beta_1&=&\frac{34}{3}N_c^2-\left(\frac{20}{3}N_c+4\, C_R\right){\rm T}_R\, n_f.
\end{eqnarray}
In the above equations, $C_R$ is the value of the quadratic Casimir operator in the color representation $R$, and ${\rm T}_R=C_R\, {\rm dim}_R/(N^2_c-1)$, where ${\rm dim}_R$ is the dimension of the representation $R$. A general method to compute $C_R$ can be found for example in~\cite{cas}. 

The crudest way of modeling a QGP would be to assume that it is an ideal gas made of gluons (Bose-Einstein distribution) and of $N_f$ quark flavors (Fermi-Dirac distribution), all assumed to be massless. Therefore, the dispersion relation is $\varepsilon(\bm k)=|\bm k|$. In that case, assuming that each quark flavor is in the same color representation $R$ and has a nonzero chemical potential $\mu$, the pressure would be given by the Stefan-Boltzmann-type law
\begin{eqnarray}\label{psb}
	p_{SB}&=&\frac{T^4}{45 \pi^2}\left\{(N^2_c-1)\pi^4\right.\nonumber\\
	&&\left.-90\left[{\rm Li}_4(-{\rm
e}^{-\frac{\mu}{T}})+{\rm Li}_4(-{\rm 
e}^{\frac{\mu}{T}})\right]{\dim}_R\, N_f\right\},
\end{eqnarray}
where ${\rm Li}_s(x)$ is the polylogarithm function of order $s$. However, if (\ref{psb}) was the QGP pressure $p$, the ratio $p/T^4$ should be constant in disagreement with all the lattice results concerning the QGP EoS. Strictly speaking, it has been shown that this last ratio is lower than the Stefan-Boltzmann case at any $T\leq 5\, T_c$, while reaching $p\approx p_{SB}$ would demand huge and probably phenomenologically irrelevant temperatures: around $(10^7,10^8)\times T_c$ following~\cite{limit}. The necessity of going beyond the Stefann-Boltzmann limit has appeared soon after the first lattice computation of the EoS in pure glue SU(2) theory~\cite{su2}. In the phenomenological studies~\cite{qp1,qp2}, it has been shown that an agreement between the lattice EoS and the one computed assuming that QCD above $T_c$ consists in an ideal gas of quasiparticles (quarks and gluons) can be reached only if the quark and gluon masses are temperature-dependent. Those thermal masses have to be fitted on the lattice data if one wants an agreement in the whole temperature range studied by lattice calculations, typically $(1.0,4.0)\times T_c$. 

Instead of introducing \textit{ad hoc} thermal masses, it is relevant to recall that perturbative techniques, or HTL resummation methods, allow to compute the thermal quark and gluon masses at large temperatures from diagrams of self-energy and vacuum polarisation-type, see for example~\cite{wel,mth1,mth2,mth3}. It appears from those studies that the thermal gluon and (anti)quark square masses are given by 
\begin{eqnarray}\label{mg}
m^2_g&=&(N_c+{\rm T}_R\, n_f)\frac{g^2}{6}T^2+\mu^2\frac{g^2}{2\pi^2}{\rm T}_R\, n_f,\nonumber\\
m^2_q&=&C_R\frac{g^2}{4}\left(T^2+\frac{\mu^2}{\pi^2}\right),
\end{eqnarray}
where a nonzero chemical potential $\mu$ is associated to the light quark flavors. Notice that in the following, $N_f$ will denote the total number of quark flavors, while $n_f$ is the number of light quarks. The ${\rm T}_R$ factor is often absent from the formulas found in the literature since it is simply equal to $1/2$ for fermions in the fundamental representation. 

The formulas~(\ref{mg}) are valid at the leading order in $g$, thus one has to be aware that our model may only be applied at temperatures well above $T_c$. An estimation of next-to-leading order corrections using HTL techniques has been made in~\cite{mth3}: The corrective terms $\delta m^2_g=-N_c\, g^2Tm_g/\sqrt 2\pi$ and $\delta m^2_q=-C_R\, g^2Tm_g/\sqrt 2\pi$ have been found. For the sake of simplicity, we nevertheless choose to keep the formulas~(\ref{mg}) in our formalism, as done in most of the quasiparticle studies. A possible justification is that such corrections mainly weakens the increase of the leading-order thermal masses with $T$. Provided that the model is not applied within a too large temperature range, as we will do, that feature can simply be mimicked in first approximation by a larger value of $\Lambda$, our free parameter.
 
\subsection{Thermodynamical properties}
The grand-canonical partition function ${\cal Z}$ of an ideal quantum gas containing quarks and gluons whose thermal masses are given by~(\ref{mg}) can be obtained using standard statistical mechanics. The pressures associated to each species can be computed thanks to the standard relation $p=T\ln{\cal Z}/V$, $V$ being the volume of the system. In units where $\hbar=c=k_B=1$, the gluon pressure reads (see \textit{e.g.}~\cite{wal})
\begin{equation}\label{pg}
	p_g=-2(N_c^2-1)\, \frac{T}{2\pi^2}\int^\infty_0dk\, k^2 \ln\left(1-{\rm
e}^{-\frac{\sqrt{k^2+m_g^2}}{T}}\right),
\end{equation}
while the contribution to the pressure of one quark flavor with chemical potential $\mu$ is given by
\begin{equation}\label{pq}
	p_q(\bar m_q)=2\ {\dim}_R\, \frac{T}{2\pi^2}\int^\infty_0dk\, k^2
\ln\left(1+{\rm e}^{-\frac{\sqrt{k^2+m_q^2+\bar m^2_q}-\mu}{T}}\right),
\end{equation}
the only change occuring at the level of the chemical potential. 
The degeneracy factor in the above expression includes the spin and color degrees of freedom. We remark that a possibly nonzero bare quark mass, $\bar m_q$, has been quadratically added to the thermal quark mass -- thermal effects come indeed as a modification of the bare quark propagator. The corresponding antiquark pressure is given by
\begin{equation}\label{pqb}
	p_{\bar q}(\bar m_q)=2\ {\dim}_R\, \frac{T}{2\pi^2}\int^\infty_0dk\, k^2
\ln\left(1+{\rm e}^{-\frac{\sqrt{k^2+m_q^2+\bar m^2_q}+\mu}{T}}\right).
\end{equation}
In general, our results will be normalized to the Stefan-Boltzmann pressure~(\ref{psb}) so that they remain finite at large $N_c$. Notice that $p_{SB}$ given by~(\ref{psb}) is nothing but the sum $p_g+N_f(p_q+p_{\bar q})$ in which $m_g=m_q=\bar m_q=0$. 

In the following we will mostly focus on the trace anomaly, which is defined as $\Delta=e-3p$ and is a particularly relevant observable since it is a direct measure of the nonideal character of the deconfined medium. It can be computed from the pressure by using the thermodynamical relation $\Delta=T^5\partial_T(p/T^4)$, or equivalently
\begin{equation}\label{delta}
\frac{\Delta}{p_{SB}}=T\, \partial_T\left(\frac{p}{p_{SB}}\right).
\end{equation}

It has been previously shown that such a quasiparticle picture is able to reproduce some lattice data from $T_c$ to higher temperatures, provided that the $T$-dependence of $g^2$ is modified, typically by replacing $\ln(T/\lambda)$ in $g^2$ by some function $\ln(h(T/\Lambda))$, with $h(x)\propto x$ at large $x$, see for example~\cite{levai,qpmodel1}. In the present work, we will adopt a different point of view. We think indeed that it is not worth modifying the above perturbative results in order to obtain a model in agreement with lattice QCD in the range $(1.0,2.0)\times T_c$. The reason is that, at those temperatures, the QGP is far from being an ideal gas, as shown by the maximal value of $\Delta/T^4$, systematically reached around 1.2 $T_c$ in lattice computations. This means that, in that range, not only the behavior of $g^2$ should be modified, but mostly the framework itself, which needs to go beyond the ideal-gas assumption. That is why we will apply the above formalism as it is, but where it is the most likely to be trusted, that is at temperatures well above $T_c$, typically above $2\, T_c$. The arbitrary is then considerably reduced.

\section{The SU$(N_c)$ gluon plasma}\label{glupla}

The simplest case to which the present approach can be applied is the gluon plasma, whose EoS has been first computed in SU(2)~\cite{su2} and then in SU(3) lattice QCD in~\cite{boyd95}, while more accurate results have recently been obtained in~\cite{panero} using gauge groups ranging from SU(3) to SU(8). 

\begin{figure}[t]
\begin{center}
\includegraphics*[width=8cm]{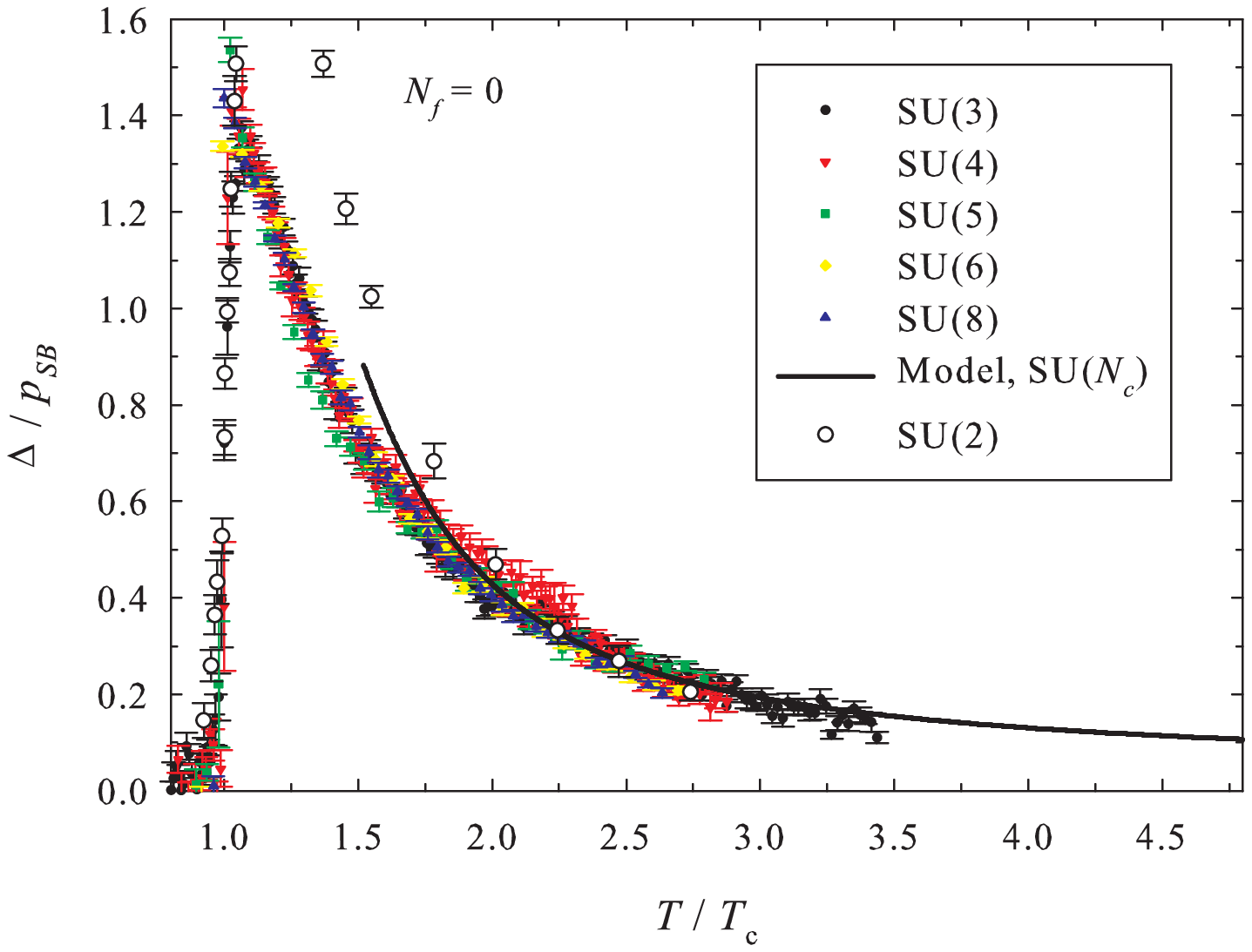} 
        \caption{(Color online) Pure glue trace anomaly, normalized to the Stefan-Boltzmann pressure~(\ref{psb}) and computed by using~(\ref{delta}) with $p=p_g$ given by~(\ref{pg}) (solid line). The result, independent of $N_c$ and computed with $\Lambda/T_c=0.88$, is compared to the corresponding lattice data taken from~\cite{panero} for the gauge groups SU(3) to SU(8) (full symbols), and from~\cite{su2} for SU(2) (empty symbols).}
\label{fig1}
\end{center}
\end{figure}

In the pure glue case, $N_f=n_f=0$ and $p=p_g$. Then, it can be observed from Eq.~(\ref{mg}) that the thermal gluon mass is independent of $N_c$ since $m_g^2\propto N_c\, g^2$. This last term is constant with respect to $N_c$ since $g^2=\lambda/N_c$ from (\ref{g2}); $\lambda$ is actually the 't~Hooft coupling. Consequently, $p_g$ is schematically equal to $(N^2_c-1)\, f(T)$, leading to ratios $p/p_{SB}$ and $\Delta/p_{SB}$ which are independent of $N_c$. That feature of the gluon plasma has indeed been observed in~\cite{panero}, where the EoS for the different gauge groups considered are found to be compatible up to the error bars (see also Fig.~\ref{fig1}). Notice that the above discussion is based on the leading-order formulation of $m_g$. In principle, higher-order terms might introduce other dependences in $N_c$; typically one could expect $m_g$ to be constant up to terms in $1/N_c^2$ as it is generally the case with gluonic observables. Such subleading corrections cannot be studied within our framework, where we focus on the dominant behavior of the EoS. We thus stress that, in the rest of this paper, the large-$N_c$ behavior of the thermodynamical quantities is meant to be that at dominant order in $N_c$, \textit{i.e.} our conclusions are valid up to subleading corrections in $N_c$.  

The general framework presented in the previous section, when particularized to the present case, describes an ideal gluon gas, and thus may be applied at temperatures where the interactions between gluons are not dominant with respect to their kinetic energy. Looking at the lattice results of~\cite{panero} in Fig.~\ref{fig1}, it appears that the gluon plasma is maximally far from an ideal gas around $1.2\, T_c$. Then $\Delta/T^4$ quickly decreases, but our approach should clearly not be applied at temperatures lower than typically $2\, T_c$, where $\Delta/T^4$ becomes well smaller than its maximal value. The coherence of that discussion would require the interaction potential between two gluons above 2 $T_c$ to be significantly smaller than around $T_c$. Informations about that potential can be deduced from lattice results concerning the free energy between static quarks: The free energy between two gluons should indeed be the same as the one between a quark-antiquark pair up to an overall color factor in virtue of the Casimir scaling~\cite{kacz1}. The gluons being light particles, they should be especially sensitive to the large-distance part of the interaction potential, and it appears from lattice calculations that the asymptotic value of the free energy between a quark and an antiquark at 2.0 $T_c$ is less than two times the one at 1.2 $T_c$~\cite{kacz2}. So it seems reasonable to trust our minimal quasiparticle model mostly well above $T_c$, where the color interactions are significantly weakened. 

The only parameter having to be fitted is the energy scale $\Lambda$, present in $g^2$. That scale is usually interpreted as a typical energy scale for the confinement. Since the temperature $T$ is used as an average energy for the system under study, it can be guessed that $\Lambda/T_c=O(1)$. Indeed, the deconfinement can intuitively be expected to occur when the average energy of the thermal fluctuations is able to ``compete" with that of the confinement, thus when $T$ becomes similar to $\Lambda$. Here $\Lambda$ is fitted to the SU(3) trace anomaly computed in~\cite{panero} in the range $(2.0,3.5)\times T_c$, that is the case for which a maximal number of data is available. In accordance with our guess, we obtain $\Lambda=0.88\, T_c$, a value that leads to an excellent agreement with the lattice data as shown in Fig.~\ref{fig1}. We have plotted our model's results up to 1.5 $T_c$ also, and it appears that the model begins to miss the data at such low temperatures, as expected. Notice that, if the lattice data are compatible with our model for all $N_c$ at high temperatures, the SU(2) lattice data are clearly distinct from the other values of $N_c$ in the range $(1.0,2.0)\times T_c$. Two interpretations of that fact can be done, not excluding each other. First, those data, coming from~\cite{su2}, are more than twenty years old. The lattice used was clearly smaller than the most recent calculations~\cite{panero}, and it is possible that what we observe is actually a lattice artefact. Second, the SU(2) data begin to differ significantly when we have argued that neglecting the interactions between gluons becomes irrelevant. Those interactions should induce corrections, probably in inverse powers of $1/N_c^2$ as it is usually the case for gluonic observables, that can be significant when $N_c=2$. Only a new lattice calculation of the SU(2) EoS could clarify that issue. 

An extension of the present framework to temperatures near $T_c$ would demand the inclusion of gluon-gluon interactions within a non-ideal Bose gas framework. A first step in that direction can be found for example in~\cite{cass,gardim}. However, such a question will not be addressed in the present work, which is an attempt to find a model describing most of the currently known lattice data with as few parameters as possible, namely $\Lambda$, $T_c$ and the quark masses. Concerning the interactions nevertheless, it is worth mentioning that two gluons can be found in the following color channels: 

\begin{eqnarray}\label{ytab}
\hspace{-1.1cm}N_c-1\left\{\begin{array}{l}  \yng(2,1) \\ \ \, \vdots \\ \yng(1)\end{array}	\right.\otimes \begin{array}{l}  \yng(2,1) \\ \ \, \vdots \\ \yng(1)\end{array}&=&\bullet\oplus  \begin{array}{l}  \yng(2,1)\\ \ \, \vdots \\ \yng(1)\end{array}\oplus \begin{array}{l}  \yng(4,2)\\ \ \, \vdots\quad\! \vdots \\ \yng(2)\end{array}\oplus \begin{array}{l}  \yng(3,3)\\ \ \, \vdots\quad\!\vdots \\ \yng(2)\end{array}\oplus \begin{array}{l}  \yng(2,1)\\ \ \, \vdots \\ \yng(1)\end{array}	\nonumber\\
&&\oplus N_c-2\left\{\begin{array}{l}  \yng(3,1,1) \\ \ \, \vdots \\ \yng(1)\end{array}	\right.\oplus \begin{array}{l}  \yng(2,2,1) \\ \ \, \vdots \\ \yng(1)\end{array}	,
\end{eqnarray}

\noindent where all the Young diagrams of the first (second) line have $N_c-1$ ($N_c-2$) rows. Notice that the last Young diagram is present only if $N_c>3$, and that only the first three diagrams exist at $N_c=2$. Any corresponding potential of one-gluon-exchange form, typically a Yukawa potential, would be proportional to the color factor $(C_{R_g}-2N_c)g^2/2$, $R_g$ being the color representation of the gluon pair. Moreover, it should involve the Debye mass, linked to the thermal gluon mass. Recalling that $g^2=\lambda/N_c$, the color factors corresponding to the Young diagrams~(\ref{ytab}) read (divided by $\lambda$) $-1$, $-1/2$, $1/N_c$, 0, $-1/2$, 0, and $-1/N_c$ respectively. Since the thermal gluon mass is constant with $N_c$ at leading order, this means that the dominant gluon-gluon interactions are independent of $N_c$. So, from a quasiparticle picture, one expects the pure glue EoS to be constant with $N_c$ in the whole range $T\geq T_c$ at the dominant order, as observed first in~\cite{panero} and confirmed in~\cite{gupta}. 

We mention for completeness that the thermodynamical properties of pure gauge QCD has been recently computed in HTL perturbation theory at three loops~\cite{3loop}; the results compare very favorably with lattice QCD down to $(2,3)\times T_c$ and the independence of the results with respect to $N_c$ is readily checked. Also, nonperturbative effects like a contribution coming from the gluon condensate could be taken into account, as suggested for example in~\cite{casto,gogo,hiet}.    

\section{Adding quark flavors}\label{addqf}

\subsection{Vanishing chemical potential}
So far we have shown that our simple quasiparticle approach is able to reproduce the lattice EoS of a gluon plasma well above $T_c$ for any gauge group provided that the parameter $\Lambda$ is fitted to the data. The next step is now to add the quark flavors to our formalism and see whether the corresponding lattice data can be reproduced or not in the same temperature range. We choose to keep the value of $\Lambda$ previously fitted to the pure glue case so that the arbitrary of the model is reduced, while the quark masses are taken from the PDG~\cite{pdg}. The only needed extra parameter is the QCD critical temperature. Lattice simulations generally find $T_c$ to be located in the interval $(150,200)$~MeV when light quark flavors are present~\cite{tcd1,tcd2,tcd3}. Here we take a quite typical value of 180~MeV, but our results would not be significantly affected using any $T_c$ located in the above interval. The parameters that will be used in the rest of this paper are 
\begin{eqnarray}\label{params}
\Lambda &=&0.88\, T_c,\quad	T_c=180\, {\rm MeV}, \nonumber\\
\bar m_u&=&\bar m_d=0\, {\rm MeV},\ \bar m_s=150\, {\rm MeV},\nonumber\\
 \bar m_c&=&1270\, {\rm MeV}, \ \bar m_b=4200\, {\rm MeV},
 \end{eqnarray}
where the $u$ and $d$ quark masses, given by 2.5 and 5 MeV respectively in the PDG, have been taken equal to zero. In the lattice calculations~\cite{detar,cheng,baza}, to which we will compare our results, the strange quark mass has nearly its physical value, $\hat m_s\approx\bar m_s$ (the hatted symbols denote the lattice masses), while the $u$ and $d$ masses are such that $\hat m_u=\hat m_d=\hat m_s/10$, that is a larger value than the PDG ones. Moreover, in~\cite{detar,detar0}, the heavy quark masses are such that $\hat m_c/\hat m_s=10$ and $\hat m_b/\hat m_s=38$, that is a charm mass compatible with the experimental one, while the lattice bottom mass is quite heavy, around 5700~MeV. We have checked that the curves we obtain are nearly indistinguishable if the $u$ and $d$ quark masses are allowed to vary from 0 to 20 MeV, \textit{i.e.} values close to the experimental and lattice ones respectively.  

\begin{figure}[t]
\begin{center}
\includegraphics*[width=8cm]{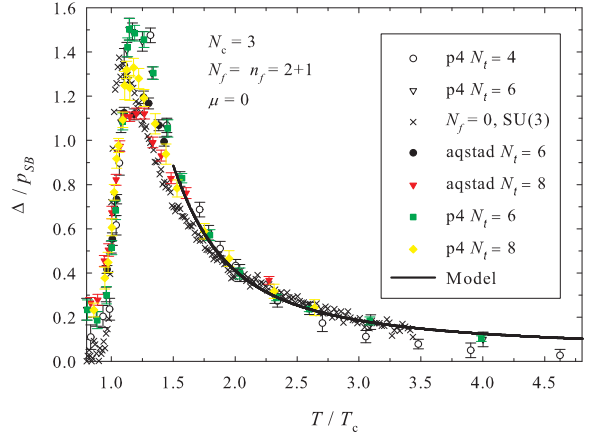} 
        \caption{(Color online) Trace anomaly of a SU(3) quark-gluon plasma with $2+1$ light quark flavors at zero chemical potential, normalized to the Stefan-Boltzmann pressure~(\ref{psb}) and computed by using~(\ref{delta}) with $p=p_{lq}$ given by~(\ref{plq}) (solid line). Parameters~(\ref{params}) are used. The result is compared to the corresponding lattice data taken from~\cite{cheng} for $N_t=4$, 6 (empty symbols), and from~\cite{baza} for $N_t=6$, 8 (full symbols). The data of both works are indistinguishable for $N_t=6$. Pure gauge results for the group SU(3), taken from~\cite{panero}, are plotted for comparison (crosses).}
\label{fig2}
\end{center}
\end{figure}

Most of the current effort in lattice QCD has logically been devoted to the study of a SU(3) QGP with $N_f=n_f=2+1$ light quark flavors and zero chemical potential, whose pressure reads in our framework
\begin{equation}\label{plq}
p_{lq}=p_g+2\, p_q(0)+2\, p_{\bar q}(0)+p_q(\bar m_s)+p_{\bar q}(\bar m_s),
\end{equation}
with $N_c=3$ and $N_f=n_f=2+1$.
The corresponding trace anomaly, computed thanks to Eq.~(\ref{delta}) in which $p_{lq}$ is used, compares very well, when normalized to $p_{SB}$, with the available lattice data from the most recent computations, involving lattices with a larger number of temporal slices $N_t=6$ and 8 (see Fig.~\ref{fig2}). Notice the following numerical coincidence: When they are expressed in units of $T_c$ and normalized to the Stefan-Boltzmann pressure, those data are very close to the ones of the pure glue case. 

\begin{figure}[t]
\begin{center}
\includegraphics*[width=8cm]{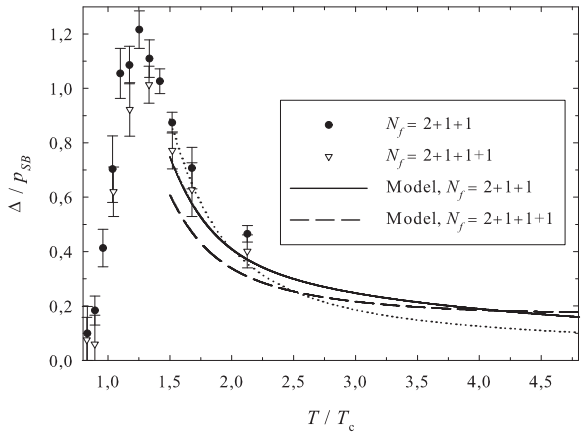} 
        \caption{Trace anomaly of a SU(3) quark-gluon plasma with $2+1$ light quark flavors plus charm (solid line) and bottom quarks (dashed line) at zero chemical potential, normalized to the Stefan-Boltzmann pressure~(\ref{psb}) and computed by using~(\ref{delta}) with $p=p_{hq}$ given by~(\ref{phq}). The trace anomaly with only $2+1$ light quark flavors is plotted for comparison (dotted line). Parameters~(\ref{params}) are used. The results are compared to the corresponding lattice data taken from~\cite{detar} (symbols).}
\label{fig3}
\end{center}
\end{figure}

Because of the statistical suppression in ${\rm e}^{-m/T}$ for a given particle species [see Eq.~(\ref{pq})], the dominant contribution to the EoS is expected to come from the lightest quarks. However, at high enough temperatures, the ratio $m/T$ can become small even with heavy quarks, although the decrease is slower than for light quarks. That is why the contribution of deconfined heavy quarks should be taken into account in more realistic approaches of the quark-gluon plasma at high temperatures. The charm and bottom quark contribution to the EoS has, to our knowledge, only been investigated on the lattice in~\cite{detar,detar0}. These recent results can be compared to our model's prediction, reading  
\begin{eqnarray}\label{phq}
p_{hq}&=&p_{lq}+p_q(\bar m_c)+p_{\bar q}(\bar m_c)\nonumber\\
&&\quad {\rm for}\ N_f=2+1+1, \nonumber\\
&=&p_{lq}+p_q(\bar m_c)+p_{\bar q}(\bar m_c)+p_q(\bar m_b)+p_{\bar q}(\bar m_b)\nonumber\\
&&\quad {\rm for}\ N_f=2+1+1+1,
\end{eqnarray}
with $N_c=3$ and $n_f=2+1$ in both cases. The maximal considered temperature is 5 $T_c\approx$ 900 MeV: The heavy quark loops can be neglected at such a typical energy, justifying our choice for $n_f$. As shown in Fig.~\ref{fig3}, no lattice data are currently known above 2 $T_c$ but our results seem to extrapolate coherently the available ones. Remark that the heavy quark contribution significantly modifies the asymptotic behavior of the results obtained with only light quarks, as illustrated in Fig.~\ref{fig3}. Notice that using the lattice $b$ mass of 5700~MeV instead of 4200~MeV only slightly modifies the curve obtained at $N_t=2+1+1+1$. We have not plotted it in Fig.~\ref{fig3} for the sake of clarity.  

\subsection{Nonzero chemical potential}

In view of an eventual comparison of quasiparticle models to experimental results, the use of a nonzero chemical potential is particularly relevant. The most recent lattice calculations deal with a nonzero constant ratio $\mu/T$ and $2+1$ light quark flavors~\cite{detar,detar0m,detar2}. Since the error bars concerning the change in trace anomaly due to a nonzero chemical potential are quite large in these works, we prefer to focus on an other observable for which the uncertainty in smaller: the change in pressure, defined in our model as
\begin{equation}\label{dpm}
\Delta p=\left.p_{lq}\right|_{\mu}-\left.p_{lq}\right|_{\mu=0}.	
\end{equation}
Notice that, when $\mu\neq0$, the Stefan-Boltzmann pressure~(\ref{psb}) also changes by an amount
\begin{equation}\label{dpsb}
	\Delta p_{SB}=\left.p_{SB}\right|_{\mu}-\left.p_{SB}\right|_{\mu=0}.
\end{equation}

\begin{figure}[t]
\begin{center}
\includegraphics*[width=8cm]{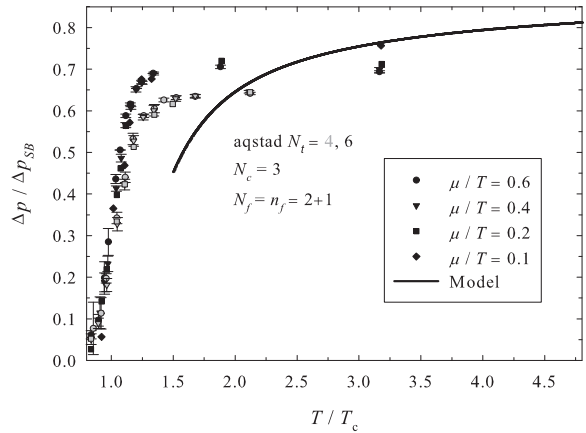} 
       \caption{Change in pressure of a quark-gluon plasma with $2+1$ light quark flavors due to a nonzero chemical potential, normalized to the change in Stefan-Boltzmann pressure~(\ref{dpsb}) and computed by using~(\ref{dpm}) with $p=p_{lq}$ given by~(\ref{plq}) (solid lines). Parameters~(\ref{params}) are used. The results are compared to the corresponding lattice data taken from~\cite{detar} (symbols). Notice that the curves obtained at different $\mu/T$ are indistinguishable.}
\label{fig4}
\end{center}
\end{figure}

In a previous lattice study~\cite{csikor}, where the EoS is studied for $2+1$ light quark flavors ($N_t=4$) at nonzero constant $\mu$ between 100 and 530~MeV. Among other results, the observation is made that the ratio $\Delta p/\Delta p_{SB}$ is constant, up to the error bars, for any chemical potential in the studied range. Following our model, those ratios differ by 3\% at 2 $T_c$ provided that $\mu$ is taken between 100 and 530 MeV; that difference moreover quickly tends to zero as $T$ increases. This confirms that, as first pointed out in~\cite{csikor}, the dependence in $\mu$ of $\Delta p$ and $\Delta p_{SB}$ factorizes in the same way, thus independently of the quark masses. As for the case where heavy quarks are added, only a few data are currently known beyond 2 $T_c$. However, a glance at Fig.~\ref{fig4} shows that the ratios $\Delta p/\Delta p_{SB}$, obtained from the lattice results of~\cite{detar}, are indeed independent of $\mu/T$ (at $N_t$ fixed), in agreement with our results. Furthermore, the present approach seems to agree with the asymptotic behavior of $\Delta p/\Delta p_{SB}$ suggested by lattice QCD. We remark that the computed curves are logically incompatible with the lattice data near $T_c$, that is where we think that our approach can certainly not be trusted anymore.    

\section{Large $N_c$ limits}\label{lnclimits}
\subsection{The different cases}\label{cases}
Despite its simplicity, the present model is defined for an arbitrary gauge group SU($N_c$) and quark representation $R$. Consequently, it allows to check the accuracy of various large-$N_c$ limits in reproducing the EoS in the QCD case, that will denote hereafter an SU(3) theory with $2+1$ light quarks in the fundamental representation. The only criterion to formulate a large-$N_c$ generalization of QCD is obviously that it should reduce to usual QCD when $N_c=3$; then results can be extrapolated to an infinite number of colors and the ``closeness" of that limit with respect to $N_c=3$ can be studied. We mainly find three different cases of interest:
\begin{itemize}
	\item 't Hooft limit: In that limit, which has been first proposed in the literature~\cite{hoof,witt}, the number of colors can become arbitrarily large while the 't Hooft coupling $\lambda=g^2 N_c$ remains constant for any $N_c$. The quarks are still in the fundamental representation, with a finite value for $N_f$. Thus, in that limit, the gluon contribution to the EoS scales as $N_c^2$ while the quark one scales as $N_c$ and can be neglected at the dominant order. 
	\item Veneziano limit: That case is similar to the 't Hooft limit, excepted that the number of quark flavors is allowed to become infinite, \textit{i.e.} $N_f=O(N_c)$~\cite{vene}. This implies that the internal quark loops in Feynman diagrams are no longer suppressed at large $N_c$. In our case, the consequence of that limit is that the quark contribution to the EoS scales as $N_c^2$ instead of $N_c$. We will take explicitly $N_f=n_f=N_c$, which leads to three light quark flavors as required for $N_c=3$.
	\item QCD$_{{\rm AS}}$ limit: When $N_c=3$, the fundamental and conjugate representations are isomorphic. One can then in principle define a quark to be in the representation $R=\yng(1,1)\equiv A_2$\, ; this is just a redefinition of the terms ``color" and ``anticolor". At large $N_c$ however, the fundamental and two-index antisymmetric representations are no longer equivalent. We call QCD$_{{\rm AS}}$ the limit in which $N_f$ remains finite, but with quarks in the $A_2$-representation. In large-$N_c$ QCD$_{{\rm AS}}$ the quark loops are not suppressed. Moreover, this case has been shown to be equivalent to a large-$N_c$ gauge theory with quarks in the adjoint representation~\cite{qcdas}. This leads in particular to an appealing duality between QCD$_{{\rm AS}}$ with $N_f=n_f=2$ and ${\cal N}=1$ SUSY Yang-Mills; we will come back to that point later.  
\end{itemize}

We mention for completeness the Corrigan-Ramond limit~\cite{corr}, which has been the first attempt to include quarks in higher representations. In that limit, $N_f$ remains finite and the quark flavors are separated into two color representations, either the fundamental one or the ($N_c-2$)-index antisymmetric one, which are both equal at $N_c=3$. This allows in particular to represent baryons as three-quark state for any $N_c$, but this limit will not be considered in the present paper, where we have chosen to keep all the quarks in the same color representation.   

\subsection{EoS at large $N_c$}
\begin{figure}[t]
\begin{center}
\includegraphics*[width=8cm]{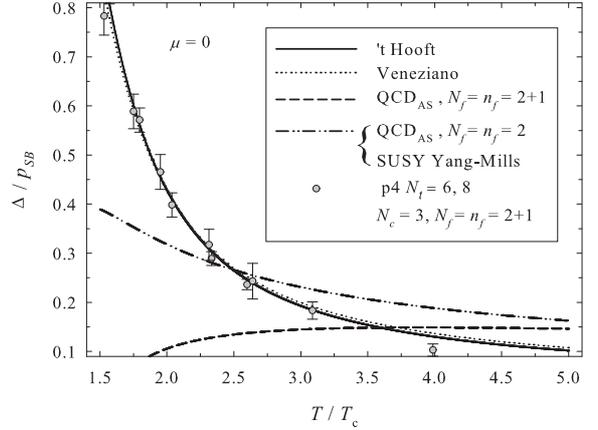} 
 \caption{Trace anomaly of a quark-gluon plasma with light quark flavors and zero chemical potential, normalized to the Stefan-Boltzmann pressure~(\ref{psb}), in the different large-$N_c$ limits defined in Sec.~\ref{cases} (lines). Parameters~(\ref{params}) are used. Lattice data taken from~\cite{baza}, corresponding to a SU(3) quark-gluon plasma with $2+1$ light quark flavors, have been added for comparison (circles).}
\label{fig5}
\end{center}
\end{figure}

We are now in position to study the behavior of the EoS in the different large-$N_c$ limits that we have proposed. The results obtained for the trace anomaly of a QGP with only light quark flavors are given in Fig.~\ref{fig5}. Several observations can be made. First of all, the 't Hooft and Veneziano limits are both excellent approximations of the QCD case, and reproduce the lattice data accurately as well. Then, the QCD$_{{\rm AS}}$ limit with $N_f=n_f=2+1$ fails to reproduce QCD, and has even a very different qualitative behavior. The reason can be understood by studying the general structure of the underlying theory, as done in~\cite{sanni}. Following the conclusions of this last work, based on the study of the two-loop $\beta$-function, the gauge theory underlying the QCD$_{{\rm AS}}$ limit is conformal for $\frac{83}{20}<n_f<\frac{11}{2}$, and looses its asymptotic freedom beyond that value. But, already for $n_f>\frac{17}{8}=2.125$, it exhibits a Banks-Zaks infrared fixed point. Such a feature, not shared by QCD, is thus present when $N_f=n_f=2+1$; we think that it explains why the physical behavior of the normalized trace anomaly is so different in that case than for QCD. If $N_f=n_f=2$ however, no fixed point is present, and  it can be observed in Fig.~\ref{fig5} that, although being in poor quantitative agreement, the qualitative behavior of the trace anomaly in that limit is at least in agreement with the QCD case. It is worth saying also that a Banks-Zaks infrared fixed point appears in the Veneziano limit if $n_f/N_c\geq34/13$~\cite{sanni}; this is not the case here since we have taken $N_f=n_f=N_c$. 

\begin{figure}[t]
\begin{center}
\includegraphics*[width=8cm]{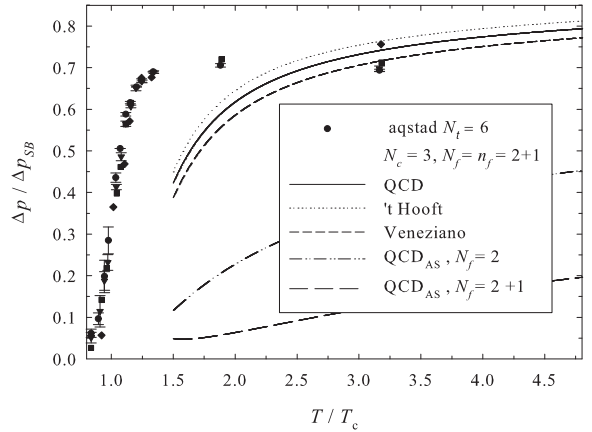} 
 \caption{Change in pressure of a quark-gluon plasma with $2+1$ light quark flavors due to a nonzero chemical potential, normalized to the change in Stefan-Boltzmann pressure~(\ref{dpsb}), in the different large-$N_c$ limits defined in Sec.~\ref{lnclimits} (lines). Parameters~(\ref{params}) are used. Lattice data taken from~\cite{detar}, corresponding to a SU(3) quark-gluon plasma with 2+1 light quark flavors, have been added for comparison (circles).}
\label{fig6}
\end{center}
\end{figure}

It is also of interest to compute the change in pressure due to a nonzero chemical potential in the different limits of interest: Results are plotted in Fig.~\ref{fig6}. As expected from the previous discussion, the QCD$_{{\rm AS}}$ limits are inaccurate with respect to the QCD case. However, the differences between the 't Hooft and Veneziano limits are more visible than for the trace anomaly; the hierarchy of the results are coherent with the relative contributions from the quark loops in those limits. Consequently, the study of the EoS at nonzero chemical potential might be relevant also in checking the relative accuracy of different inequivalent large-$N_c$ limits of QCD.  

\subsection{Adjoint quarks and SUSY Yang-Mills}\label{susydef}

Among the various gauge theories built on the SU($N_c$) gauge group, a particularly interesting case is the one with one flavor of massless fermions in the adjoint representation. Indeed, such a theory is supersymmetric and it is nothing else than a ${\cal N}=1$ SUSY Yang-Mills theory~\cite{salam}, the adjoint quarks being also called the gluinos. Within our framework, the EoS of that supersymmetric theory can be computed by taking the following pressure as an input:
\begin{equation}\label{psusy}
p_{SUSY}=p_g+p_q(0),
\end{equation}
with the quarks in the adjoint representation and $\mu=0$. The observation of the equations presented in Sec.~\ref{qpmodel} reveals that the fermionic contributions to the thermal masses only depend on the factors ${\rm T}_{ R}\, n_f$ and $n_f$ dim$_{{\rm R}}$ respectively. They are equal to $N_c$ and ($N^2_c-1$) respectively in the present case and lead, as in the pure glue case, to an EoS of the form $(N^2_c-1)\, f(T)$. The ratio $\Delta/p_{SB}$ is thus predicted to be independent of $N_c$ for ${\cal N}=1$ SUSY Yang-Mills. 

Figure~\ref{fig5} clearly shows that the SUSY Yang-Mills EoS is identical to the one of the large-$N_c$ QCD$_{{\rm AS}}$ limit with $N_f=n_f=2$. Again, the observation of our formalism helps to understand that feature, since
\begin{eqnarray}
	\left.{\rm T}_{ R}\, n_f\right|_{SUSY}&=N_c,  \quad \qquad\left.{\rm T}_{ R}\, n_f\right|_{QCD_{{\rm AS}}}=\frac{N_c-2}{2}n_f,
\nonumber\\	\left.{\rm dim}_{{\rm R}}\, n_f\right|_{SUSY}&=N_c^2-1, \ \left.{\rm dim}_{{\rm R}}\, n_f\right|_{QCD_{{\rm AS}}}=\frac{N_c(N_c-1)}{2}n_f.\nonumber\\
\end{eqnarray}
All those factors are equal at large $N_c$ if $n_f=2$, leading to identical thermodynamical properties. This provides another explicit check of the equivalence shown in~\cite{qcdas}. Two remarks remain to be done. First, at large $N_c$, a theory with 2 light quarks in the symmetric representation $\yng(2)$ (such as used in technicolor models~\cite{sanni}) would have the same EoS as SUSY Yang-Mills. Second, to our knowledge, the lattice EoS of the ${\cal N}=1$ SUSY Yang-Mills theory remains to be computed. 

\subsection{Comments on heavy mesons}
Not only free heavy quarks are of physical relevance to describe the QGP, but also heavy mesons. It is indeed widely believed that the color interactions are strong enough at $T\gtrsim T_c$  to bind the heavy quarks into mesons. More information can be found for example in the review~\cite{satz}. For completeness, let us make some comments about the possible large-$N_c$ generalisations of heavy meson models.

The static $q\bar q$ potential is the basic ingredient of a standard heavy meson model. It can be either the $q\bar q$ free energy, $F_1$, or internal energy, $U_1=F_1-T\, \partial_T F_1$  (see \textit{e.g.}~\cite{shuryak} for a discussion concerning that choice). Assuming as in Sec.~\ref{glupla} that the free energy above $T_c$ is mainly of one-gluon-exchange form, as suggested by recent analytical perturbative calculations~\cite{potT} and by lattice computations~\cite{free}, one has $F_1\propto -C_R\, \lambda/N_c$ for a $q\bar q$ pair in a color singlet. Comparing the $C_R$-factors with the fundamental and $A_2$ representations, one finds
\begin{equation}
	C_{\yng(1)}=\frac{N^2_c-1}{2N_c}, \quad 	C_{\yng(1,1)}=\frac{(N_c+1)(N_c-2)}{N_c}.
\end{equation}	
Those factors are equal if $N_c=3$, but at large-$N_c$ they differ by a factor 2, leading to a much more attractive potential in the QCD$_{{\rm AS}}$ limit. In the case of heavy mesons also, this last limit is thus expected to exhibit a behavior quite far from QCD. 

\section{Conclusions}\label{conclu}

We have proposed a minimal quasiparticle approach to describe the deconfined regime of a generic QCD-like theory, with SU($N_c$) gauge group and quarks in an arbitrary representation. The model is minimal in the sense that we have chosen to use the thermal quark and gluon masses computed by using perturbative methods in finite-temperature QCD, without extra modifications. The parameters are the quark masses, taken from the PDG, the critical temperature, for which a standard value is used, and the $\Lambda$-parameter of the running coupling constant. It can be fitted on the pure glue case and then taken as an input in all the other cases. Its optimal value is found to agree with the naive expectation $\Lambda/T_c=O(1)$.

Despite our model's simplicity, we have first shown that it is able to satisfactorily reproduce the equations of state computed in lattice QCD well above $T_c$ (typically above 2\, $T_c$) either in the pure glue case, or with light quark flavors at zero chemical potential. When heavy quark flavors or a nonzero chemical potential are taken into account, no lattice data are known at high temperatures. However, since our model may be working well in those cases also, it is worth mentioning that it seems to extrapolate coherently the existing lattice data. We hope that lattice data at nonzero chemical potential and higher temperatures will be soon available in order to check our model's predictions.

Then, we have particularized our framework to the case of three inequivalent large-$N_c$ limits of QCD: 't Hooft, Veneziano, and QCD$_{{\rm AS}}$. It can be observed that, in the pure glue sector, all those limits acts the same since in that case no quark flavors are present and the thermodynamical quantities normalized to the Stefann-Botlzmann pressure are explicitly independent of $N_c$ at the dominant order. The universality of the results with respect to $N_c$ has been observed on the lattice~\cite{panero} for $N_c\geq 3$ up to the error bars. That universality seems to hold even for $N_c=2$ at $T\gtrsim 2\, T_c$, but not at lower temperatures. Since the SU(2) EoS is more than twenty years old~\cite{su2}, it might be of interest to make new computations at $N_c=2$ with larger lattices to see whether the observed low-temperature difference is physical (and thus a consequence of the corrections due to gluon-gluon interactions) or a lattice artefact. 

When light quark flavors are added, the three considered large-$N_c$ limits are inequivalent. It appears that the Veneziano limit, followed by the 't Hooft limit, reproduces at best the QCD equation of state with light quarks and zero chemical potential. It also appears that the change in thermodynamical properties due to a nonzero chemical potential is an observable which is particularly sensitive to the considered large-$N_c$ limit. The QCD$_{{\rm AS}}$ limit, thus when quarks are in the two-index antisymmetric representation, has the same qualitative behavior as QCD when only two light quark flavors are present, but a poor quantitative agreement. With three light quark flavors however, even the qualitative agreement is lost. We think that this change of qualitative behavior occurs because the corresponding gauge theory develops a Banks-Zaks infrared fixed point, absent in QCD~\cite{sanni}. 

Finally, we have explicitly checked that the equation of state in the QCD$_{{\rm AS}}$ limit with two massless quark flavors is equal to the one of the ${\cal N}=1$ SUSY Yang-Mills theory, \textit{i.e.} one massless quark flavor in the adjoint representation. That last supersymmetric theory begins to be studied on the lattice at $T=0$~\cite{susyYM}; we hope that calculations at $T>0$ will soon be available to confirm our model's predictions. 

It is worth mentioning that such a quasiparticle picture should not, in our opinion, be applied at lower temperatures unless color interactions are explicitly added to the formalism. The maximal value of $\Delta/T^4$, systematically occurring at $T\approx 1.2\, T_c$, is a clear indication that an ideal-gas approach is questionable near $T_c$ and that an interacting quantum gas formalism should be used. We leave that issue for future works. 

\section*{Acknowledgments} 
F. B. thanks the F.R.S.-FNRS for financial support. The authors thank C. Semay for valuable comments about the manuscript.    

\end{document}